\documentclass[12pt,preprint]{aastex}  % for FASTTRACK submission!
\slugcomment{Submitted to ApJ Letters}

\newcommand{\xtej}{XTE J1751--305}
\newcommand{\saxj}{SAX J1808.4--3658}
\newcommand{\rxte}{{\it RXTE}}
\newcommand{\chandra}{{\it Chandra}}
\newcommand{\sax}{{\it BeppoSAX}}
\newcommand{\xmm}{{\it XMM-Newton}}

\begin{document}

\title{Discovery of a Second Millisecond Accreting Pulsar: XTE J1751--305}
\author{C. B. Markwardt,\altaffilmark{1,2}
        J. H. Swank,\altaffilmark{2}
	T. E. Strohmayer,\altaffilmark{2}
	J. J. M. in 't Zand,\altaffilmark{3,4}
        F. E. Marshall\altaffilmark{2}}

\altaffiltext{1}{Department of Astronomy, University of Maryland, 
	College Park, MD 20742; craigm@lheamail.gsfc.nasa.gov}
\altaffiltext{2}{Laboratory for High Energy Astrophysics, 
        Mail Code 662, NASA Goddard Space Flight Center, Greenbelt, MD 20771}
\altaffiltext{3}{Astronomical Institute, Utrecht University, P.O. Box 80000,
                 NL - 3508 TA Utrecht, the Netherlands}
\altaffiltext{4}{SRON National Institute for Space Research, Sorbonnelaan 2,
                 NL - 3585 CA Utrecht, the Netherlands}

\begin{abstract}
We report the discovery by the \rxte\ PCA of a second transient
accreting millisecond pulsar, \xtej, during regular monitoring
observations of the galactic bulge region.  The pulsar has a spin
frequency of 435 Hz, making it one of the fastest pulsars.  The
pulsations contain the signature of orbital Doppler modulation, which
implies an orbital period of 42 minutes, the shortest orbital period
of any known radio or X-ray millisecond pulsar.  The mass function,
$f_x = (1.278 \pm 0.003)\times 10^{-6} M_\sun$, yields a minimum mass
for the companion of between 0.013 and 0.017 $M_\sun$, depending on
the mass of the neutron star.  No eclipses were detected.  A previous
X-ray outburst in June, 1998, was discovered in archival All-Sky
Monitor data.  Assuming mass transfer in this binary system is driven
by gravitational radiation, we constrain the orbital inclination to be
in the range 30\arcdeg--85\arcdeg, and the companion mass to be
0.013--0.035 $M_\sun$.  The companion is most likely a heated helium
dwarf.  We also present results from the \chandra\ HRC-S observations
which provide the best known position of \xtej.
\end{abstract}


\section{Introduction}

Accreting neutron stars in low mass X-ray binaries (LMXBs) are
conventionally thought to be the progenitors of millisecond or
``recycled'' radio pulsars \citep{alpar82}.  Firm evidence supporting
this theory remained elusive until the launch of NASA's {\it Rossi
X-ray Timing Explorer\/} (\rxte) in December, 1995.  The discovery of
300--600 Hz nearly coherent oscillations during thermonuclear X-ray
bursts \citep[e.g.,][]{stroh97} was a first solid indicator that
neutron stars in LMXBs rotate rapidly.  This was followed by the
discovery in April, 1998, of the first accreting millisecond pulsar,
\saxj, with a spin period of 2.5 ms and orbital period of 2.1 hr
\citep{wijnands-vdk98,chakmorgan98}.  This discovery convincingly
established a link between accreting neutron stars and recycled
pulsars.  The presence of quasi-periodic oscillations in the range
300--1300 Hz in many LMXBs has also been used to infer rapid neutron
star spin \citep[e.g.,][]{vdk00-khz}.

Binary evolution models are becoming more sophisticated
\citep{podsiad02}, but still involve significant assumptions about
mass transfer and the effects of magnetic fields.  \citet{kulkarni88}
have questioned whether the birthrate of LMXBs can account for the
number of millisecond radio pulsars.  On the other hand, there have
been speculations that there should be a significant number of
low-luminosity transient LMXBs in the galaxy \citep{heise99,king00},
whose mass transfer and binary separation are driven primarily by the
emission of gravitational radiation.  Although the discovery of \saxj\
provided convincing evidence that recycled pulsars can form in LMXBs,
it is difficult to draw inferences on binary and stellar evolution
based on a single case.
% Clearly, finding more
% examples of actively accreting millisecond pulsars would assist in
% population modeling.

In this paper we report the discovery of an accreting millisecond
pulsar, \xtej, which was discovered by \rxte\ in regular monitoring of
the galactic center region.  This is the fastest known accreting
pulsar and the second of its kind to be found.  Recently, a third
pulsar XTE J0929$-$314, was discovered
\citep{remillard02,galloway02-psr}.  Interestingly, all three systems
have very low mass companions.  In \S\ref{sec:xte} and
\S\ref{sec:chandra}, we present the discovery by \rxte, and results of
a short \chandra\ observation to determine the source position.  In
\S\ref{sec:timing} we develop a pulsar timing solution, and in
\S\ref{sec:spectral} we present basic spectral results.
\S\ref{sec:disc} contains a discussion of the binary system
properties.  In this paper we focus on the pulsar timing properties,
and defer more detailed analyses of other issues to future work.

% These new findings are demonstrating that a little recognized
% population of low luminosity transient LMXBs does exist and many of
% them apparently harbor rapidly rotating neutron stars.

\section{\rxte\ Observations\label{sec:xte}}

\xtej\ was discovered using the \rxte\ Proportional Counter Array
(PCA) in a monitoring program of the galactic bulge region
\citep{swankmark01}.  The PCA instrument has an effective area of
$\sim$6500 cm$^2$, and is sensitive to 2--60 keV X-rays within a
collimated field of view, which has a triangular profile and a full
width at half-maximum (FWHM) of 1\arcdeg.  A region of approximately
250 square degrees around the galactic center region has been scanned
by the PCA twice weekly since February 1999, except for several months
when sun constraints interfere.  The nominal $1\sigma$ sensitivity of
the scans to variations is approximately 0.5--1 mCrab, but the
sensitivity is degraded somewhat within a few degrees of the galactic
center, where source confusion becomes important.

% Individual source count rates are modulated by the PCA collimator as
% they pass into and out of the field of view.  The resulting light
% curves are fitted to a model of known sources, convolved with the
% collimator response function.

In a bulge scan on 2002 April 3.62, we detected a source whose
identification was not previously known \citep{mark02-disc}.
Subsequently, detections were made and positions were determined by
the \sax\ WFC and \xmm\ \citep{intzand02-pos,ehle02}.  All of the
positions are essentially consistent with a previously unknown source
situated 2.1 degrees from the galactic center.

% \footnote{The uncertainty of the PCA determination
% is itself somewhat uncertain, due to the crowded X-ray field.}

% \begin{deluxetable}{llllc}
% \tablecaption{Position Determinations of \xtej\label{tab:position}}
% \tablehead{\colhead{Description}&\colhead{R.A. (J2000)\tablenotemark{a}}&
% 	   \colhead{Dec. (J2000)}&\colhead{$\sigma$}&\colhead{Ref.}}
% \startdata
% PCA Scan         &  17 51 30    & $-$30 30      & 5\arcmin & 1 \\
% BeppoSax WFC     &  17 51 16    & $-$30 37 30   & 1\farcm2 & 2 \\
% XMM EPIC MOS2    &  17 51 13.5  & $-$30 37 22   & 10\arcsec & 3 \\
% %\chandra\ HRC-S &  17 51 13.52 & $-$30 37 22.9 & 0\farcs6 & 4 \\
% % Following data have been corrected for aspect offset
% \chandra\ HRC-S  &  17 51 13.49 & $-$30 37 23.4 & 0\farcs6 & 4 \\
% \enddata
% \tablenotetext{a}{Values for R.A. are Hours, Minutes, Seconds; 
%   for Declination they are Degrees, Arcminutes, Arcseconds}
% \tablerefs{1---\citet{mark02-disc}, 2---\citet{intzand02-pos}, 
%            3---\citet{ehle02}, 4---\citet{mark02-pos}} 
% \end{deluxetable}

Follow-up \rxte\ pointed observations continued from April 4.6--30.9
on an essentially daily basis, for a total good exposure time of 398
ks.  X-ray pulsations at a frequency of $\approx$435 Hz were discovered
in the first 200 seconds of the pointed observation on April 4.6, with
a semi-amplitude of $\simeq 5\%$ \citep{mark02-disc}.  This makes
\xtej\ only the second accreting millisecond pulsar to be discovered.

The PCA can be configured in a variety of data modes to optimize the
usage of telemetry and science return.  The primary PCA data modes for
timing were {\tt GoodXenon} and {\tt E\_125us\_64M\_0\_1s}, which
provide 1 and 125 $\mu$s temporal resolutions, respectively.
Spectroscopy was performed using the {\tt Standard2} mode.  For light
curves and spectroscopy, the PCU0 detector was excluded, because of a
missing propane layer which increases the background level.  For high
resolution timing, all PCU detectors were used.

% The primary PCA data mode for timing on the first two days was ``{\tt
% GoodXenon},'' which provides full spectral and temporal ($1 \mu$s)
% information.  The remainder of the data were taken primarily in ``{\tt
% E\_125us\_64M\_0\_1s}'' mode, which has a coarser energy resolution
% and 125 $\mu$s time resolution, but has a lower telemetry usage.

\section{\chandra\ Observations \label{sec:chandra}}

We obtained \chandra\ HRC-S target of opportunity observations of
\xtej\ on April 10.7 \citep{mark02-pos}.  The HRC-S is a micro-channel
plate instrument sensitive to 0.1--10 keV X-rays, which can be placed
at the \chandra\ focal plane.  Due to a problem in the assignment of
event times in the on-board electronics, the HRC must be operated in a
special ``Imaging'' mode with background reducing filters disabled.
Event timestamps are then corrected in the ground software.  The total
exposure was 2969 s.  A correction for spacecraft bore sight offset
was also applied.  A source of $\sim$16000 counts was clearly detected
at a position consistent with those of \sax\ and \xmm\ (Table
\ref{tab:timing}).  The nominal position error for \chandra\ is
dominated by uncertainties in the spacecraft aspect
solution\footnote{\tt http://cxc.harvard.edu/cal/ASPECT/celmon/}
\citep{aldcroft00}, whose 90\% confidence uncertainty is quoted in
Table \ref{tab:timing}.  No other sources were detected in the HRC-S
field of view, which might have helped to refine the position further.
To date, this is the best known position of \xtej.

% \footnote{See ``Check for aspect offset and fix''
% \chandra\ analysis thread, May 2, 2002 ({\tt
% http://cxc.harvard.edu/\-cal/\-ASPECT/\-fix\_offset/\-fix\_offset.cgi}).}

% The instrument was configured in a special ``Imaging'' mode in order
% to mitigate the excess background counting rate from exceeding the
% telemetry bandwidth limit.  

% It is in principle possible to refine the position even further if one
% or more known sources are detected in the field, and a frame tie can
% be formed to a high precision catalog.  We searched for other sources
% using the XIMAGE version 4.1 software, but none were found above a
% $4\sigma$ threshold.  Therefore we assign the nominal \chandra\ aspect

\section{X-ray Timing \label{sec:timing}}

\begin{figure}
\plotone{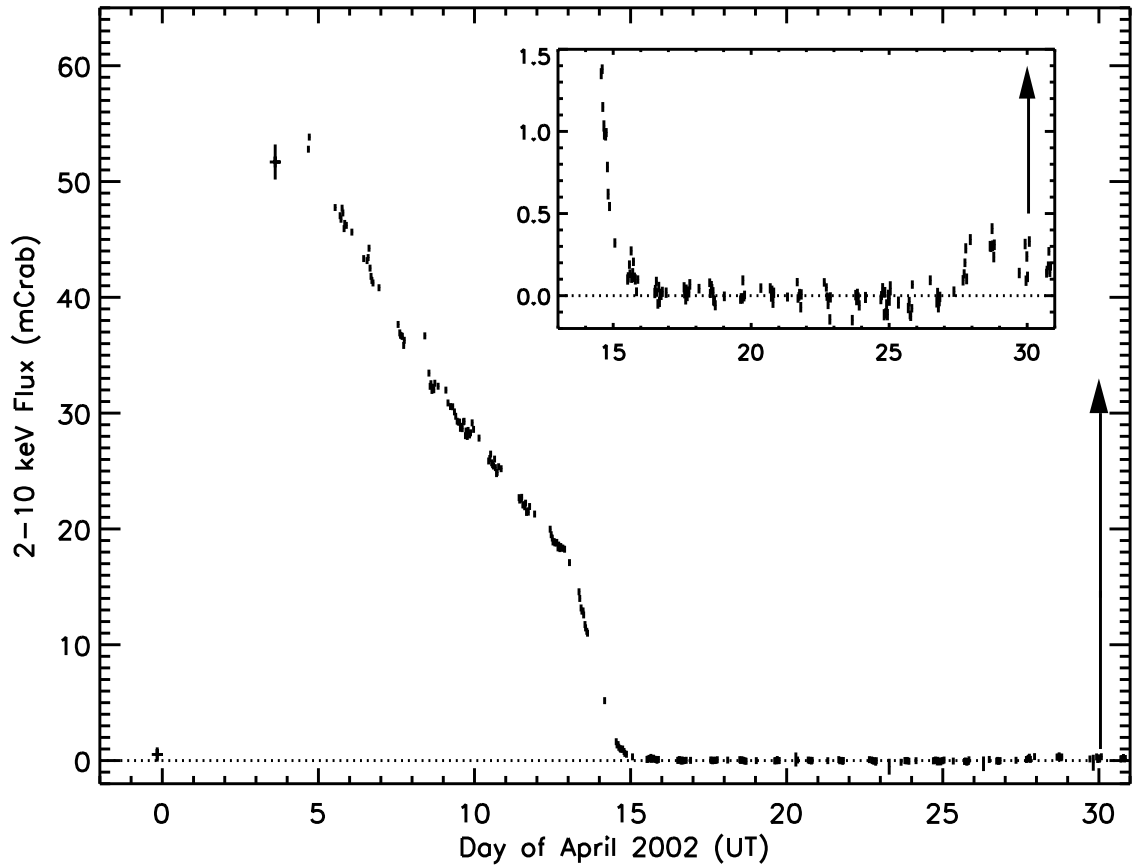}
% \centerline{\psfig{file=fig-ltlc.ps,angle=+90,width=\textwidth}}
\caption{\rxte\ PCA light curve of \xtej\ in the 2--10 keV band from
galactic bulge monitoring (crosses) and pointings (vertical bars; 1600
s bins).  The inset shows an expanded vertical scale.  The arrow
indicates the time of an X-ray burst.  Note that 1 mCrab (2--10 keV) =
$2.42\times 10^{-11}$ erg s$^{-1}$ cm$^{-2}$ = 2.27 ct s$^{-1}$
PCU$^{-1}$.
\label{fig:ltlc}}
\end{figure}

An X-ray light curve of both scanning and pointed data is shown in
Figure~\ref{fig:ltlc}.  Throughout this paper, particle background
subtraction was performed using the ``CMVLE'' model available from the
\rxte\ Guest Observer Facility.  In addition to particle backgrounds,
there is also a significant astrophysical background component due to
nearby sources and galactic diffuse emission, which affects the
baseline determination.  The mean quiescent flux levels in the scan
and pointing data (2.4 ct s$^{-1}$ and 5.63 ct s$^{-1}$ per PCU
respectively) were subtracted.

Figure~\ref{fig:ltlc} shows that the X-ray peak flux probably occurred
around April 4.5, and thereafter the light curve exhibits a nearly
exponential decay, $e^{-t/\tau}$, where $\tau = 7.1\pm 0.1$ d (compare
to the same fit for \saxj, which yields $\tau = 14\pm 1$ d).  A sudden
turn-off occurs around April 13, which is very similar to the turn-off
of \saxj.  The rise time is less than four days.  The figure also
shows that there was a detection of a small outburst of $\sim 0.3$
mCrab around April 27--30, and an X-ray burst on April 30.9.  Given
the large PCA field of view, the source identification must be done
with care.  A preliminary analysis of the burst indicates that it is
not from \xtej, but that the low-level flux may be.

%  which is accurate to within $\pm 3 \mu s$ of UTC
% (Markwardt, in preparation)

\begin{figure}
\plotone{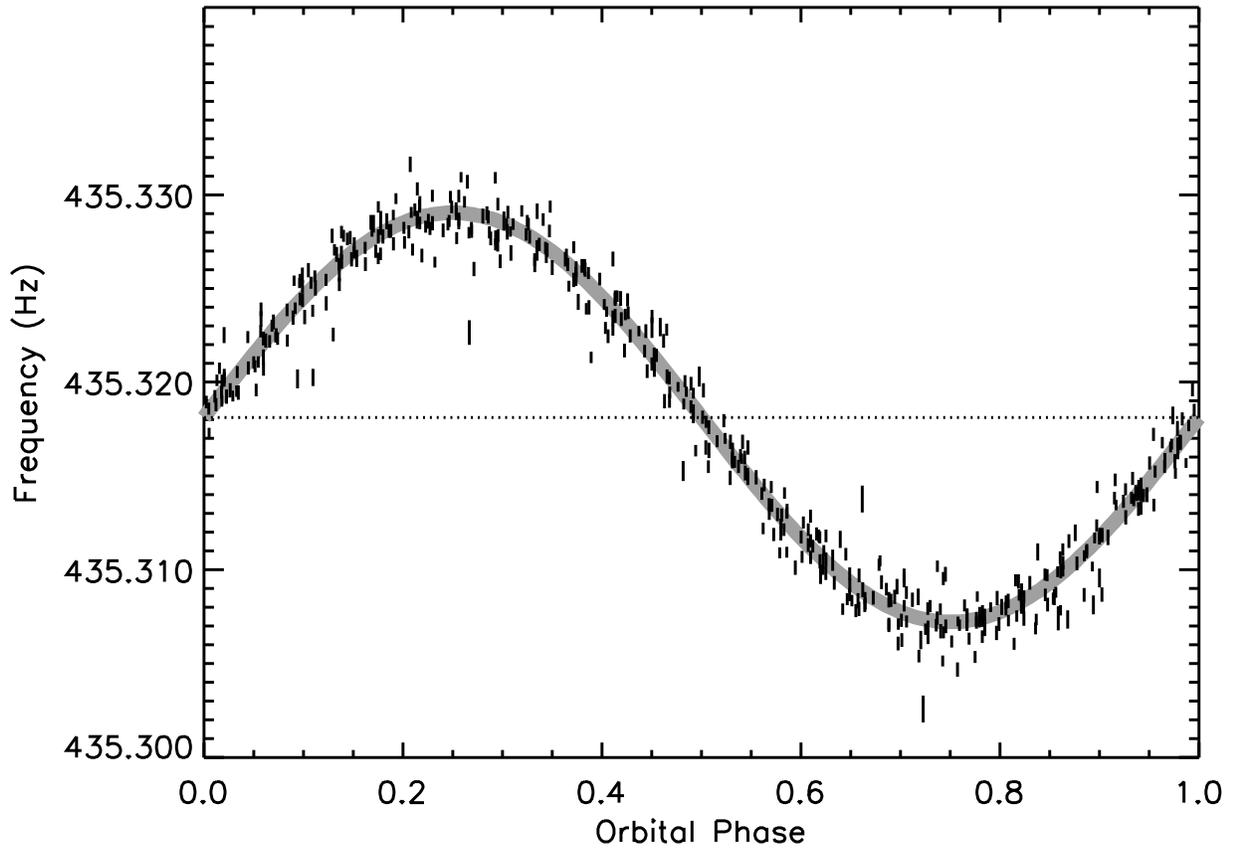}
% \centerline{\psfig{file=fig-orbfit.ps,angle=+90,width=\textwidth}}
\caption{Measured pulse frequency as a function of orbital phase,
folded on a trial period of $P_b = 2545.35$ s, for data from April
4.6--8.7.  The best fit sinusoid model is shown (thick line).
\label{fig:orbfit}}
\end{figure}

When a more complete set of data were available, a sinusoidal
variation in the centroid frequency became apparent.  To
systematically investigate this effect, we first corrected the X-ray
event arrival times at the satellite to the solar system barycenter
using the JPL planetary ephemeris DE405, the definitive \rxte\ orbit
ephemeris, and the best known source position.  We also applied a fine
clock correction, which has a magnitude of 53--78 $\mu$s.  Next, we
divided the data into 100 s segments and computed the Rayleigh or
$Z_1^2$ statistic \citep{buccheri83}, which is defined as
\begin{equation}
Z_1^2 = \left(\sum_j \cos \phi(t_j)\right)^2 + 
        \left(\sum_j \sin \phi(t_j)\right)^2
\end{equation}
where $\phi(t_j)$ is the pulse phase of a photon which arrives at time
$t_j$, according to a model of pulse phase evolution.  For the initial
investigation, $\phi(t) = 2\pi ft$ for a grid of constant frequencies
$f$ around 435 Hz, and thus the $Z_1^2$ statistic is essentially the
Fourier power spectrum in a narrow bandpass.  Each $Z_1^2$ transform
peak was fitted by a Gaussian function.  The pulsation frequencies for
data from April 4.6--8.7, folded on a trial orbital period, are shown
in Figure~\ref{fig:orbfit}, and clearly establish the orbital
frequency modulation.

% \begin{equation}
% Z_n^2 = \sum_{i=1}^{n}\left[\left(\sum_j \cos n\phi(t_j)\right)^2 + 
%                             \left(\sum_j \sin n\phi(t_j)\right)^2\right]
% \end{equation}

The timing model was refined by constructing a single pulse phase
connected solution over the data span from April 4.6--14.0.  Events
from PCA channels 5--40 from all enabled PCUs (energies of 2.5--16.9
keV) were coherently summed to form a single $Z_1^2$ statistic.  For
the phase model, we translated to IDL a version of the BNRYBT binary
pulsar model \citep{bt76} from the program TEMPO \citep{taylor89}.
The parameters of the binary were varied iteratively in order to
achieve the global maximum value of the $Z_1^2$ statistic.  Table
\ref{tab:timing} contains the optimal binary parameters.  After this,
we separated the data set into smaller (10 ks) segments and examined
the timing residuals by computing the cosine and sine terms of the
$Z_1^2$ statistic separately and converting to a time delay.  The
r.m.s. residual was $\sim 30$ $\mu$s.

% Maximization of
% the $Z^2$ is equivalent to least squares fitting of times of arrival,
% but without the need to form a template or perform cross correlation.

% In this case the X-ray pulse profile is highly sinusoidal (see below),
% so the use of the $Z_1^2$ statistic is justified, as opposed to higher
% order statistics.  

We estimated the parameter confidence intervals by employing a Monte
Carlo estimation procedure.  An ideal pulsar light curve model was
constructed with similar orbital and spin properties to \xtej.  An
ensemble of 100 light curve realizations was drawn from this model
using Poisson statistics.  Each realization was then converted to
events, and a $Z_1^2$ optimization was performed.  For all parameters
of the model, the sample variances corresponded very closely to a
region within $\Delta Z_1^2 = 1$ of the peak value.  Therefore, we
associate this region with the $1\sigma$ single-parameter confidence
region.  For the actual X-ray data we stepped each parameter of
interest through a fixed grid while allowing the other parameters to
be optimized.  Confidence intervals were set conservatively using
$\Delta Z_1^2 = 9$, corresponding to $\sim 3\sigma$ uncertainties, but
we quote enough precision to recover the $1\sigma$ errors.

% While this may appear to be a novel result, similar
% precisions have been obtained using an odds ratio test
% \citep{zavlin00,gl96}.

Using our best timing solution, we constructed an overall folded X-ray
pulse profile for the time range April 4.0--14.0.  The pulsed fraction
is defined as $p_f = \sqrt{2(Z^2-2)/N_{ev}}$, which is then rescaled
to account for background.  The mean pulsed fraction in the
fundamental was 4.41\%$\pm$0.02\%, but slowly decreases throughout the
outburst from 5.3\% to 3.9\%.  After the light curve turn-off of April
13, the pulsations are still detected at 3--4\% on April 14.  The mean
amplitude of the first harmonic was 0.12\%$\pm$0.02\%, increasing from
$<0.12\%$ on April 4.6--6.5 (95\% confidence), to 0.27\%$\pm$0.05\% on
April 11.5--15.0.  The amplitude of the second harmonic was $<
0.04\%$, with no evidence of time evolution.

% The dominance of the fundamental also justifies
% the choice of $Z_1^2$ for the timing analysis.

We also examined the \chandra\ HRC-S data for pulsations.  We
barycentered the data and selected 16414 events within $8.5\arcsec$ of
the source position, and computed the $Z_1^2$ statistic using the PCA
timing solution.  We found $Z_1^2 = 27.65$ (a $\sim 5\sigma$
detection), which implies a pulsed fraction of about 5.6\%.  Cross
calibration of the HRC and PCA data is a subject of further research.

% Upon expanding to a grid search, the peak $Z_1^2\ (=27.85)$ occurred
% within 20 $\mu$Hz of the expected peak based on the PCA solution, and
% is also within the expected $1\sigma$ confidence limits.

Finally, we applied the PCA solution to the small outburst of April
28--30.9 and tested for pulsations.  While the data for April 29.9
alone indicates a detection of $Z_1^2 = 6.8$ (a significance of
$2.1\sigma$), there are no pulsations detected in the other days from
April 27.8--30.9.  Thus the overall 95\% upper limit for pulsations in
the small outburst is 5.5\%, and an unambiguous identification with
\xtej\ on the basis of pulsations is not possible.

% the overall (incoherent) signal level is only
% significant at a $1\sigma$ level, which we do not deem to be strong
% enough to warrant an identification with \xtej.

\section{Spectral Analysis \label{sec:spectral}}

We sought to determine the total broadband X-ray flux by modeling the
X-ray spectrum.  For the purposes of this paper, we chose
representative observations on April 5.6, 9.2 and 12.7, which
correspond to an approximately even sampling of the light curve decay.
PCA and HEXTE spectra were extracted using the analysis techniques
recommended by the \rxte\ Guest Observer Facility.  The PCA response
matrix was computed using PCARSP version 8.0.  The HEXTE instrument
has two clusters of detectors which are sensitive to 15--200 keV
X-rays and has field of view of $\sim 1^\circ$ (FWHM).  The collimated
detectors are physically rocked back and forth by $\pm1.5^\circ$ in
order to sample time-varying particle backgrounds.  In this analysis,
spectra from HEXTE cluster ``A'' were background subtracted using the
$-1.5^\circ$ rocking position.

% ; the spectra from the $+1.5^\circ$
% position were clearly contaminated by another source.

% We verified that once the source became undetectable in the PCA, the
% HEXTE ``source'' spectrum also converged to the background level.

The joint spectrum is well modeled by an absorbed power law with
photon index 1.7--1.9, and an exponential cut-off energy of between
100 and 200 keV.  The spectrum remains more or less constant during
the decline of the outburst, and so we estimate that approximately
$2.6\pm0.5$ times as much flux appears in the 2--200 keV band as in
the 2--10 keV band depicted in Figure~\ref{fig:ltlc}.  Using this
conversion we determine the integrated outburst fluence to be
$(2.5\pm0.5)\times 10^{-3}$~erg cm$^{-2}$ (2--200 keV).

% The neutral hydrogen absorption value ranged from
% $(1.5-2.5)\times 10^{22}$ cm$^{-2}$.

%   However, \citet{miller02-epic} has found a low
% temperature black body component ($kT_{BB} = 1.06$ keV) in XMM-Newton
% EPIC spectra of the source --- which is largely below the PCA band ---
% so we do not ascribe much physical meaning to the absorption value.

\begin{figure}
\plotone{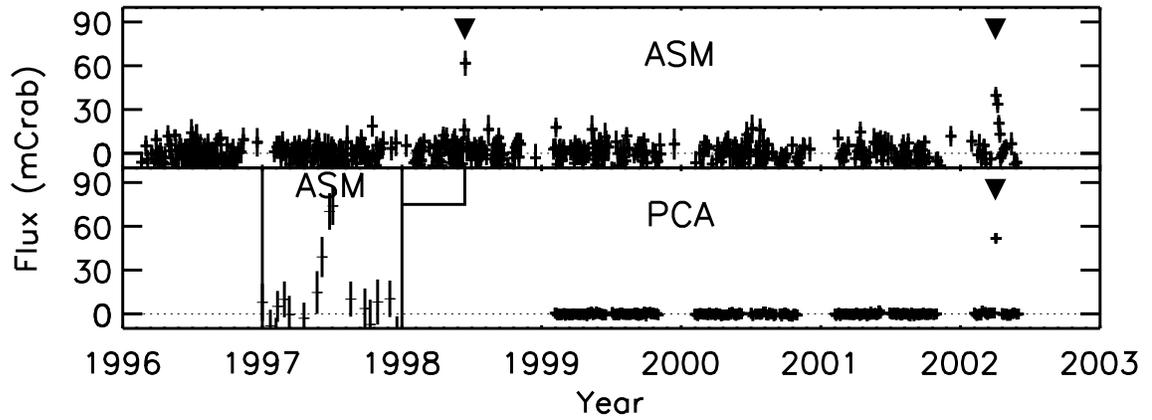}
% \centerline{\psfig{file=fig-asm.ps,angle=+90,width=\textwidth}}
\caption{\xtej\ light curves.  (top) ASM (2--12 keV); (bottom) PCA
bulge monitoring (2--10 keV).  (bottom inset) ASM Light curve from
June 6--26, 1998.  Triangles highlight the two outburst dates.  
\label{fig:asm}}
\end{figure}

We searched the \rxte\ All Sky Monitor (ASM) light curve for \xtej,
available from the MIT ASM web site.  After filtering the data to
remove points with large uncertainty ($> 2.5$ ct/s), and rebinning to
form one day averages, we detected another probable outburst in June
of 1998 (Figure \ref{fig:asm}).  The peak flux in the 2--12 keV band
was approximately $80\pm 10$ mCrab, and the duration was 3--4 days.
We also examined the PCA bulge monitoring data since February 1999,
but no outbursts exceeding 0.5 mCrab were detected.  While there are
gaps in the ASM and PCA monitoring programs where a several day
outburst would be missed, we tentatively assign $T_{\rm rec} = 3.8$ yr
as the mean recurrence time.

% The \sax\ WFC has
% accumulated over $\sim 6$ Ms of galactic center observations since
% 1996 (irregularly spaced) with no X-ray bursts or persistent emission
% detected from the source position \citep[in 't Zand 2002, private
% communication;][]{intzand01-bursters}.  

\section{Discussion \label{sec:disc}}

% M_jup = 0.00095 M_sun

The binary orbital parameters allow us to estimate the properties of
the companion.  The 42 minute orbital period immediately reveals that
\xtej\ is a highly compact binary system.  \xtej\ has the shortest
orbital period of all known millisecond pulsars (both X-ray and
radio).  The amplitude of the modulations also determines the mass
function of the pulsar shown in Table~\ref{tab:timing}, defined as
$f_x = (M_c \sin i)^3 / (M_x + M_c)^2$, where $i$ is the binary
inclination to our line of sight, and $M_x$ and $M_c$ are the masses
of the pulsar and companion respectively.  Given $M_x$, the mass
function provides the minimum possible value of $M_c$, which would
occur when viewing the binary edge-on ($i = 90^\circ$).  The minimum
mass range shown in Table~\ref{tab:timing} reflects a reasonable range
of neutron star masses between 1.4--2.0 $M_\sun$.  It is clear that
the companion of \xtej\ is in the regime of very low mass dwarfs, of
order 15 Jupiter masses.  This is a factor of $\sim$3 smaller than the
companion of \saxj, which \citet{bildchak00} speculate is a $0.05
M_\sun$ brown dwarf.

\begin{figure}
\plotone{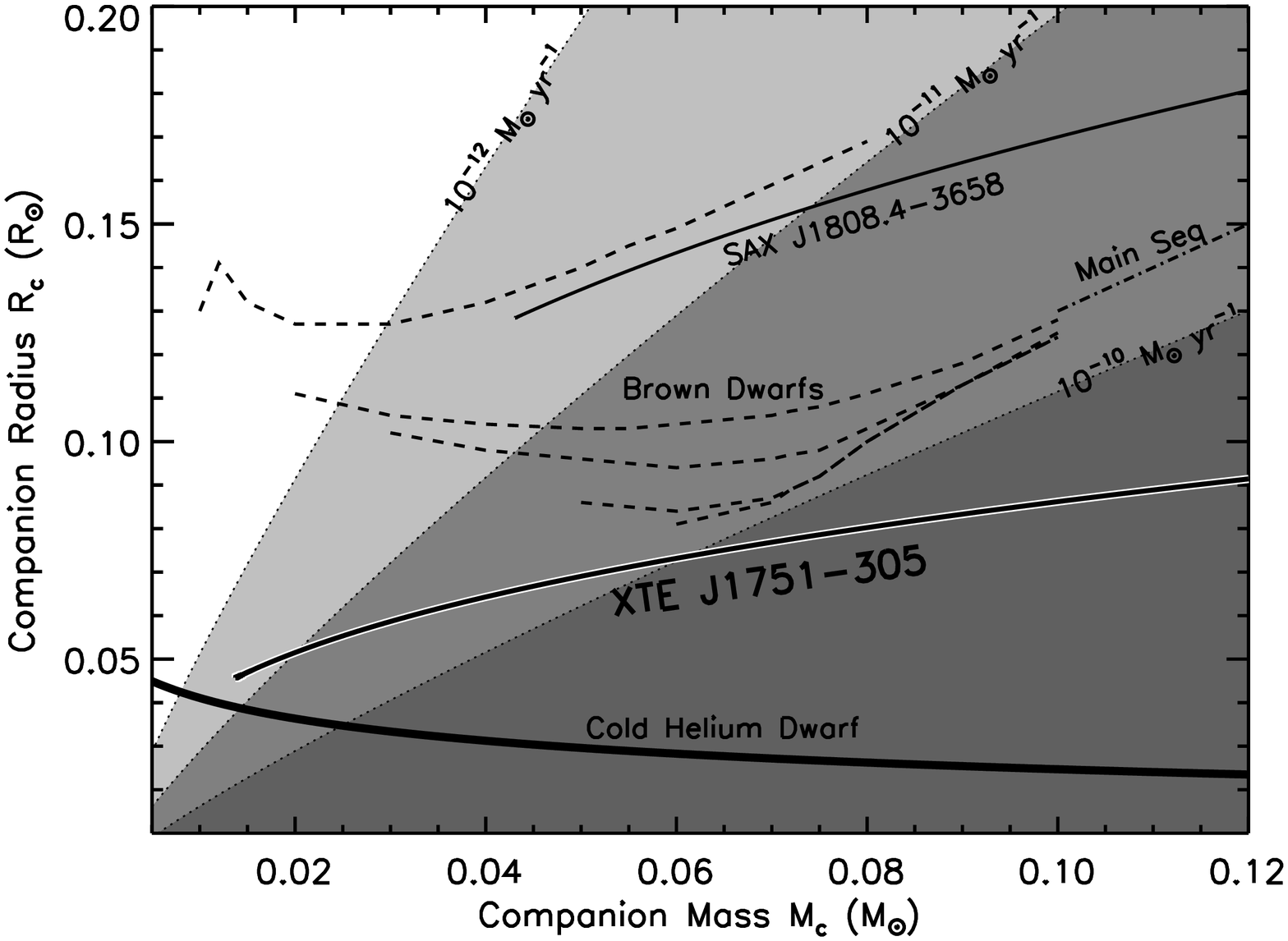}
% \centerline{\psfig{file=fig-mvr.ps,angle=+90,width=\textwidth}}
\caption{Companion mass ($M_c$) vs. radius ($R_c$) plane, showing the
Roche lobe constraints for \xtej\ (highlighted solid) and \saxj (thin
solid).  The equations of state are shown for hydrogen main sequence
(dash-dot), brown dwarfs (dashed) and a cold helium dwarf (thick
solid).  Shaded contours of represent lines of constant $\dot{M}_{GR}$
for a $1.4 M_\sun$ neutron star \citep{rappaport84}.  Brown dwarf
models are for ages of (from top to bottom) 0.1, 0.5, 1, 5 and 10
billion years.
\label{fig:mvr}}
\end{figure}

It is reasonable to assume that the companion must fill its Roche lobe
in order to transfer mass to the neutron star \citep{eggleton83}.
Combining the mass function and Roche lobe constraints results in a
curve in the $M_c$ vs. $R_c$ plane, shown in Figure~\ref{fig:mvr}.
This line can be compared to the equations of state of other types of
bodies, including hydrogen main sequence stars, brown dwarfs
\citep{chabrier00}, \saxj\ \citep{bildchak00}, and a cold helium dwarf
\citep{zapolsky69,rappaport84}.  While none of the models intersect
the trace of \xtej, the hydrogen models are clearly unlikely.  The
oldest brown dwarf model is unlikely, given the probability that
irradiation by the compact object will cause bloating
\citep{bildchak00}.  A warm helium dwarf model, which would lie above
the ``cold helium dwarf'' curve in Figure~\ref{fig:mvr}, appears to be
the most likely scenario, but there appear to be no calculations of
such a configuration in the literature.

There were no X-ray eclipses or dips detected in the PCA light curves.
We therefore put an upper limit of $i<85^\circ$.  There is also no
evidence of X-ray modulations at the binary period (at a limit of
$~\sim 0.5\%$), which might have implied propagation through a
scattering atmosphere in a near edge-on geometry \citep{bildchak00}.
On the other hand, a very low inclination is also unlikely, since that
would imply a high companion mass, and thus a large mass transfer
rate.  Following the reasoning of \citet{bildchak00}, we find that the
time-averaged mass accretion rate, based on the measured X-ray fluence
and a recurrence time of 3.8 yr, is $\dot{M}_x = 2.1\times 10^{-11}
M_\sun$ yr$^{-1} d_{10}^2 m_{1.4}^{-1} T_{3.8}^{-1}$, where $d_{10}$
is the distance in units of 10 kpc, $m_{1.4} = M_x / (1.4 M_\sun)$,
and assuming a neutron star radius of 10 km.  On the other hand, the
mass transfer in these systems is thought to be driven by
gravitational radiation \citep{king00,rappaport84}, in which case the
mass transfer rate should be $\dot{M}_{GR} = 1.2\times 10^{-11}
M_\sun$ yr$^{-1} m_c^2 m_{1.4}^{2/3}$, where $m_c = M_c / (0.0137
M_\sun$), and we have assumed the companion is nearly degenerate.  If
the two mass transfer rates are equal, we arrive at the constraint,
$\sin i = 0.74 m_{1.4}^{3/2} d_{10}^{-1}$.  For distances within 15
kpc, this constraint implies inclinations of 30\arcdeg--85\arcdeg and
companion masses in the range 0.013--0.035 $M_\sun$, for neutron star
masses between 1.4 $M_\sun$ and 2.0 $M_\sun$.

An interesting result of this constraint is that the distance is at
least 7 kpc, significantly farther than \saxj\ \citep[$d = 2.5$
kpc;][]{intzand01-1808}.  The peak luminosity of the persistent
emission would then be $\gtrsim 2\times 10^{37}$ erg s$^{-1}$, an
order of magnitude higher than that of \saxj\ \citep{cui98}.  It is
probable that \xtej\ is near the galactic center, in which case more
pulsars like it should be detectable by the PCA bulge monitoring
program over the next few years.

\acknowledgments This work was partially supported by a NASA
Astrophysics Data Program grant.

\begin{deluxetable}{lc}
\tablecaption{Timing Parameters of \xtej\label{tab:timing}}
\tablehead{\colhead{Parameter} & \colhead{Value}}
\startdata
Right ascension, $\alpha$ (J2000)                      & 
      $17^{\rm h}51^{\rm m}13\fs49(5)$\tablenotemark{a}\\
Declination, $\delta$ (J2000)                          & 
      $-30\arcdeg37\arcmin23\farcs4(6)$\tablenotemark{a}\\
Barycentric pulse frequency, $f_o$ (Hz)                & 435.317993681(12)\tablenotemark{b} \\
Pulsar frequency derivative, $|\dot f|$ (Hz s$^{-1}$)  & $<3\times 10^{-13}$\\
Projected semimajor axis, $a_x \sin i$ (lt-ms)         & 10.1134(83) \\
Binary orbital period, $P_b$ (s)                       & 2545.3414(38) \\
Epoch of mean longitude $90^\circ$, $T_{90}$           & 54118.7563591(87)\tablenotemark{c} \\
Orbital eccentricity, $e$                              & $<1.7\times 10^{-3}$\\
Pulsar mass function, $f_x$ ($10^{-6} M_\sun$)         & 1.2797(31) \\
Minimum companion mass, $M_c$ ($M_\sun$)               & 0.0137--0.0174 \\
Maximum Power, $Z_{\rm max}^2$                         & 36237 \\
\enddata
\tablenotetext{a}{Parameter was fixed; 90\% confidence limits from \chandra\
aspect uncertainty.}
\tablenotetext{b}{Uncertainties and upper limits are $3\sigma$ in last quoted digits.}
\tablenotetext{c}{Modified Julian days, referred to TDB timescale.}
\end{deluxetable}

% 1 Crab = 75 ASM ct s$^{-1}$.

\end{document}